\documentclass[onecolumn,10pt, superscriptaddress]{revtex4-2}
\usepackage[paperwidth=210mm,paperheight=297mm,centering,hmargin=4cm,vmargin=2.5cm]{geometry}
\usepackage{graphicx}
\usepackage{amsmath, amssymb, graphics, setspace}
\usepackage{bm} 
\usepackage{dcolumn}
\usepackage{color}
\usepackage{xcolor}
\usepackage{mathrsfs}
\usepackage{xfrac}
\usepackage{amsfonts}
\usepackage{varioref}
      \usepackage[skip=0.333\baselineskip]{caption}
\usepackage{subcaption}
        \usepackage{lmodern}
        \usepackage{caption}
         \usepackage{floatrow}%
         \usepackage[english]{babel}
\usepackage[utf8]{inputenc}
\usepackage{mathtools}
\usepackage{placeins}
\usepackage[T1]{fontenc}
\usepackage{lipsum}
\usepackage{csquotes}
\usepackage{siunitx}
\usepackage[export]{adjustbox} 
\usepackage{bm}
\usepackage{ragged2e}
\usepackage[pdftex, pdftitle={Article}, pdfauthor={Author},colorlinks=true, linkcolor=blue, citecolor=blue]{hyperref}
\input epsf
\usepackage{float}
\labelformat{section}{Section #1} 
\labelformat{subsection}{Section #1} 
\labelformat{subsubsection}{Section #1}
\labelformat{subsubsubsection}{Section #1}
\labelformat{equation}{Eq.~(#1)} 
\labelformat{figure}{Fig.~#1} 
\labelformat{subfigure}{Fig.~\thefigure#1} 
\labelformat{table}{Tab.~#1} 
\labelformat{appendix}{Appendix #1}

\newcommand{\LF}{\left(}
\newcommand{\RF}{\right)}
\newcommand{\LT}{\left[}
\newcommand{\RT}{\right]}
\newcommand{\Ld}{\left.}
\newcommand{\Rd}{\right.}
\newcommand{\ba}{\begin{eqnarray}}
\newcommand{\ea}{\end{eqnarray}}
\newcommand{\be}{\begin{equation}}
\newcommand{\ee}{\end{equation}}
\newcommand{\bi}{\begin{itemize}}
\newcommand{\ei}{\end{itemize}}
\newcommand{\non}{\nonumber\\}
\begin{document}
\title{\Large Constant Curvature 3-branes in 5-D f(R) Bulk}
\author{ Shafaq Gulzar Elahi
\footnote{shafaqelahi2028@u.northwestern.edu}}
\affiliation{Northwestern University, Evanston, Illinois-60201, USA}
\author{ Soumya Samrat Mandal
\footnote{mandal15@purdue.edu}}
\affiliation{Purdue University, West Lafayette, Indiana-47907, USA}
\author{ Soumitra SenGupta \footnote{tpssg@iacs.res.in}}
\affiliation{Indian Association for the Cultivation of Science (IACS),
Kolkata-700032, India}

\maketitle
\section*{Abstract}
\textit{Braneworld models remain the most promising candidates to address several important questions in low-energy particle phenomenology and cosmology. The role of the moduli field(s) and its stabilization is an integral part of this question. In this work, we show that a 5-dimensional warped braneworld model with higher curvature gravity in bulk admits de-Sitter and anti de-Sitter solutions on the branes. The remarkable feature of having a positive vacuum energy on the visible brane is the presence of a metastable minimum and a global minimum for the modulus potential. While the metastable minimum leads to a consistent cosmological model of a bouncing universe, the concomitant existence of the global minimum provides a vacuum for the modulus to roll down to stability. Further, this model is shown to be consistent with the swampland conjecture to qualify as a viable candidate in the low energy description of string landscape}

\section{Introduction}
Ever since the proposal of Kaluza and Klein, there have been extensive studies to explore the possible roles of the extra spatial dimension in our 3-dimensional universe. Various implications of the higher dimensional geometry on lower dimensional physics range from gauge hierarchy problem \cite{fine-tuning}, the origin of dark matter/ dark energy \cite{galaxy}-\cite{SG}, neutrino mass hierarchy\cite{neutrino},\cite{neutrino2}, gauge/gravity duality\cite{KS}, big bang or bouncing cosmology\cite{KK3}-\cite{bounce4}, etc. One of the generic features of all kinds of extra-dimensional models is the presence of one or more moduli fields associated with the geometry of the internal manifold. While extracting an acceptable theory for our 3+ 1 dimensional universe continues to be a big challenge, it is also very hard to propose a mechanism that points  towards  a unique solution to the geometry of the extra dimensions.
While string theory helped to screen out to some extent the nature of the internal manifold  using the swampland conjecture\cite{Ooguri:2006in}-\cite{Montefalcone:2020vlu}, however, the observational signatures of the dynamics of the moduli field in cosmological/astrophysical scenarios are expected to play a crucial role in exploring the possible signatures of extra dimensions. In this context, the braneworld models are the most important candidates to bear signatures of the extra dimensions and have been explored extensively. The simple models for braneworld physics assume the existence of one extra dimension and look for its signature on the lower dimensional brane. Although a proper resolution of the origin of hierarchical scales in our brane namely the electroweak and Planck scale continued to elude us for a long time,
the presence of extra dimension is found to offer a possible resolution to this problem. However, it was realized that such models often bring in 
new hierarchies in disguise through the scale associated with internal geometry. A typical example of this was encountered in large extra-dimensional models\cite{large},\cite{large2}-\cite{Arkani-Hamed:1998jmv}.
As an alternative, a warped geometry model such as  the Randall-Sundrum model (RS) \cite{RS1} could address different issues without introducing any hierarchical scale in the model. The model is based on a space-time which is fundamentally 5-dimensional with the background being $AdS_5$ and the extra spatial dimension $y=\phi r_c$ (where $r_c$ is the compactification radius) being angular and subject to $S_1/Z_2$ compactification. Consequently, there are two 3-branes at the orbifold fixed points, the TeV/Visible and the Planck/Hidden Brane. The metric under consideration is a non-factorizable one i.e. the 4-D part of the metric is multiplied by an exponential function of the extra-dimension.
The five-dimensional action is:
\begin{equation}
\label{(1)}
\mathcal{S}=\int{ d^5x\,\sqrt{-g}(\text{M}^3\text{R}-\Lambda_5)}-\int {d^4x \,\sqrt{-g_i}\mathcal{V}_i}
\end{equation}
where $\text{M}$ is the fundamental 5-D Planck scale (we are working in natural units), $\text{R}$ is the 5-D Ricci Scalar,$\Lambda_5$ is the bulk cosmological constant, $\cal{V}$$_i$ is the tension of the ith brane(i=hid(vis)) and $\eta_{\mu\nu}$is the 4D metric.\\
The RS solution implies a negative bulk cosmological constant  $\Lambda_5=-12\text{M}^3k^2$ where $k=\sqrt{-\Lambda_5/12M^3}$ and brane tensions $\cal{V}$${}_{\text{hid}}=-$$\cal{V}$${}_{\text{vis}}=12\text{M}^3k^2$.
The Solution is:\\
 \begin{equation} \label{metric-RS} ds^2=e^{-2k r_c|\phi|}\eta_{\mu\nu}dx^{\mu}dx^{\nu}+r_c^2d\phi^2\end{equation}
A natural extension to the RS model is to include the higher curvature terms in the action which is  motivated by the fact that the 5-D is endowed with a large cosmological constant  $\mathcal{O}$$\sim\rm  M$ which in turn ensures that no intermediate scale unlike large extra-dimensional models develop in the theory. However, the gravity action with such a large background curvature most naturally should be described higher curvature correction to Einstein's action.
In \cite{Elahi}, we considered the simplest extension to the Einstein-Hilbert Action in the bulk in the form of  $\rm f(R)=R+\alpha R^2-\Lambda_5$  since the bulk has the scalar curvature $\rm R\sim \mathcal{O}(M^2)$ and found the corresponding modification to the warp factor in a perturbative manner. We also showed that the presence of additional gravitational degrees of freedom in the bulk due to f(R) gravity could automatically generate a stabilizing potential for the modulus/radion field. Meanwhile, in an earlier work \cite{SD}, the RS model was generalized to include non-flat maximally symmetric 3-branes. This work produced a wide class of warped solutions where the RS model was just a special case when the brane cosmological constant was zero. The de-Sitter generalization was particularly relevant to address various cosmological scenarios with the accelerated expansion of the universe. One interesting outcome of considering a non-flat brane was subsequently shown \cite{Banerjee} in the context of modulus stabilization. The outcome indicated the generation  of a nonzero modulus potential where the ground state of the modulus indicated a metastable vacuum. 
The potential and the vacuum value of the modulus were proportional to the brane cosmological constant which disappears if the brane cosmological constant is set to zero as in the original RS case. It was further shown in \cite{bounce4} that the existence of such a modulus potential with a metastable ground state can lead to a transient  phantom phase resulting in a  bouncing cosmological model with a low scalar tensor ratio consistent with Planck data. However, the possibility of having an appropriate global minimum for the radion field could not be resolved in this model. 
Moreover, despite several interesting outcomes, this non-flat generalization was done with Einstein action in the bulk though the bulk continued to have a vacuum curvature of the order of the Planck scale. The presence of higher curvature terms in the action should be a more appropriate description in such a scenario. 
Motivated by our earlier work \cite{Elahi} generalizing the warped geometry model  with higher curvature  gravity with flat 3-branes, in this work, we want to analyze the effect of considering a non-zero 4-D cosmological constant in our universe i.e. the 4-D metric is no longer $\eta_{\mu\nu}$, instead, it has a constant curvature with a positive or negative brane cosmological constant leading to a de-Sitter or anti de-Sitter 3-brane. Our goal is to explore the nature 
of the modulus potential and the consequent modulus stabilization in such a scenario.
\section{The Model}
Throughout the paper, we will adapt the mostly positive convention for the metric i.e. $\eta_{\mu\nu}=(-,+,+,+)$, the Greek indices $\alpha,\beta$.. denote 4-D indices whereas the Latin indices a,b, A, B... denote the 5-D indices.\\
Consider the metric ansatz:
\be
\label{metric}
ds^2=e^{-2\text{A(y)}}\overline{g}_{\mu\nu}(x)dx^\mu dx^\nu +\text{B(y)}^2dy^2
\ee

Here $\overline{g}_{\mu\nu}$ is the 4-D metric.The non-zero Christoffel symbols for this metric are : $\Gamma^5_{55}=\text{B(y)}'/\text{B(y)}$,
$\Gamma^5_{\mu\nu}=\overline{g}_{\mu\nu}e^{-2\text{A(y)}}\text{A(y)}'/\text{B(y)}^2$ , $\Gamma^\mu_{5\nu}=-\delta^\mu_\nu \text{A(y)}'$and ${}^4\Gamma^\alpha_{\mu\nu}$. Here, prime denotes the derivative w.r.t. the extra-dimensional coordinate y. Also,
\begin{align}
   & \text{R}={}^4\mathcal{R}e^{2\text{A(y)}}+\frac{-4}{\text{B}(y)^3}\left[2 \text{A(y)}'\text{B}'(y)+\text{B}(y)\left(5 \text{A(y)}'^2-2 \text{A(y)}''\right)\right]\label{(6)}\\
    &\text{R}_{55}=-4\left[\frac{\text{A}'(y)\text{B}'(y)}{\text{B}(y)}-\text{A}''(y)+\text{A}'(y)^2\right]\label{(7)}\\
    &\text{R}_{\mu\nu}={}^4 \rm \mathcal{R}_{\mu\nu}+\frac{1}{\text{B}(y)^3}\overline{g}_{\mu\nu}e^{-2\text{A(y)}}\left[\text{B}(y)\left(\text{A}''(y)-4 \text{A}'(y)^2\right)-\text{A}'(y)\text{B}'(y)\right]\label{(8)}
\end{align}
where ${}^4\mathcal{R}$ is the 4-D scalar curvature. We will consider that the 5-D scalar curvature is only a function of the extra-dimension i.e., $\rm R=R(y)$. This is because we chose ${}^4 \mathcal{R}=4\Omega$ and ${}^4 \mathcal{R}_{\mu\nu}=\Omega \overline{g}_{\mu\nu}$, where $\Omega$ is a constant and can be both positive or negative. We will set $\rm M=1$ for simplification.
We want to explore constant curvature 3-brane solutions of RS-like space-time when the bulk action is given by:
\be
\label{action}
\mathcal{S}= \int{ d^5x\,\sqrt{-{g}}~\text{f(R)}}-\int {d^4x \,\sqrt{-g_i}\mathcal{V}_i}
\ee
where $\rm f(R)=R+\alpha R^2-\Lambda_5$. The field equation for a general f(R) action is:
\begin{equation}
\label{EOM}
\text{R}_{\text{MN}} \text{f}_\text{R}(\text{R})-\frac{1}{2} g_{\text{MN}} \text{f(R)}+g_{\text{MN}} \square  \text{f}_\text{R}(\text{R})-\nabla_{\text{M}} \nabla_{\text{N}} \text{f}_\text{R}(\text{R})=\frac{1}{2}\textbf{T}_{\text{MN}}
\end{equation}
where $\textbf{T}_{\text{MN}}$ is the energy-momentum tensor corresponding to matter of some kind.
Substituting \ref{(6)},~\ref{(7)} and \ref{(8)} in \ref{EOM}, we obtain the following equations (for the bulk):
\ba\label{bulk-eqn}
&&\rm-\frac{1}{B(y)^4}~2 \left(-3 B(y)^4 A'(y)^2+\Omega  e^{2 A(y)} B(y)^6+3 k^2 B(y)^6\right)\non
&&\rm-\frac{2 \alpha}{B(y)^4}~ \Biggl\{80 A'(y)^2 B'(y)^2+8 \Omega  e^{2 A(y)} B(y)^4 A'(y)^2+32 B(y) A'(y) \left(2 B'(y) \left(2 A'(y)^2\right.\right.\non
&&\rm\left.\left.-A''(y)\right)-A'(y) B''(y)\right)+4 B(y)^2 \left(-4 A''(y)^2+5 A'(y)^4+8 A'''(y) A'(y)\Rd\non
&&\rm\Ld-32 A'(y)^2 A''(y)\right)+4 \Omega ^2 e^{4 A(y)} B(y)^6\Biggr\}=0
\ea
\ba\label{boundary-eqn}
&&\rm -\frac{1}{B(y)^7}\LF3 B(y)^5 A''(y)-3 B(y)^4 A'(y) B'(y)-6 B(y)^5 A'(y)^2+B(y)^7 \left(\Omega  e^{2 A(y)}+6 k^2\right)\RF\non
&&\rm-\frac{1}{B(y)^7}8 \alpha  \Biggl\{\Omega  e^{2 A(y)} B(y)^4 A'(y) B'(y)+30 A'(y) B'(y)^3+10 B(y) B'(y) \left(3 B'(y) \left(2 A'(y)^2\Rd\Rd\non
&&\rm\Ld\Ld-A''(y)\right)-2 A'(y) B''(y)\right)+\Omega  e^{2 A(y)} B(y)^5 \left(A'(y)^2-A''(y)\right)+B(y)^2 \left(12 A'''(y) B'(y)+8 A''(y) B''(y)\Rd\non
&&\rm\Ld-16 A'(y)^2 B''(y)+37 A'(y)^3 B'(y)+2 A'(y) \left(B'''(y)-36 A''(y) B'(y)\right)\right)+B(y)^3 \left(-2 A''''(y)+12 A''(y)^2\Rd\non
&&\rm\Ld\Ld+5 A'(y)^4+16 A'''(y) A'(y)-37 A'(y)^2 A''(y)\right)\right)\Biggr\}=0
\ea
\subsection{Perturbative Solution}
 Consider $\alpha$ to be small such that $ \text{R}^2$ in $\text{f(R)}$ is a small correction over $\text{R}$ solution, In this backdrop, consider the following ansatz for $\text{A(y)}$ and $\text{B(y)}$:
\ba\label{ansatz}
&&\rm e^{-A(y)}= \omega ~sinh \LT ln \left(\frac{c_2}{\omega }\right)-k~y\RT+\alpha A_1(y)\\
&&\rm B(y)= 1+\alpha b_0~y
\ea

where $\omega^2=\Omega/3k^2$ for $\Omega>0$ and $\rm \omega ~\rm sinh \LT ln \left(\frac{c_2}{\omega }\right)-k~y\RT$ represents the unperturbed solution as per \cite{SD} i.e. by setting $\alpha \to 0$ in \ref{bulk-eqn}and~\ref{boundary-eqn}. The constant $c_2=\rm\sqrt{1+\sqrt{1+\omega^2}}$ is obtained by setting the boundary condition at the hidden brane $\rm e^{-A(y=0)}=1$.We have chosen to fix $\rm B(y)$ as per \cite{Elahi} and find a solution for $\rm A(y)$.\\
$\cal{O}$$(\alpha^0)$ is trivially satisfied and at $\cal{O}$$(\alpha)$, \ref{bulk-eqn} and \ref{boundary-eqn} reduce to:
\ba\label{bulk-orderalpha}
&&\rm -\frac{1}{c_2~k}~3 \Omega  e^{k y} \left(k A_1(y)-A_1(y)'\right)+9 c_2~ k e^{-k y} \left(A_1(y)'+k~ A_1(y)\right)\non
&&\rm +\frac{1}{6 c_2^2 ~k^2}\Omega ^2 e^{2 k y} \left(3 b_0 ~y+10 k^2\right)+\frac{3}{2} c_2^2 ~k^2~ e^{-2 k y} \left(3 b_0~ y+10 k^2\right)+3 b_0~ y \Omega -10 k^2 \Omega=0
\ea
\ba\label{boundary-orderalpha}
&&\rm 54 c_2^3~ k^3 ~e^{k y} \left(-A_1(y)''+2 k ~A_1(y)'+3 k^2 A_1(y)\right)-18 c_2 ~k \Omega  e^{3 k y} \left(-A_1(y)''-2 k~ A_1(y)'\Rd\non
&&\rm \Ld\rm +3 ~k^2 A_1(y)\right)+3 b_0 \left(9 c_2^4~ k^4~ (4 k y-1)+\Omega ^2 e^{4 k y} ~(4 k y+1)\right)+40 k^3 \left(\Omega  e^{2 k y}-3 c_2^2~ k^2\right)^2=0
\ea
The solution is:
\be\label{A1-solution}
\rm A_1(y)=\frac{1}{12} c_2~ k e^{-k y} \left(-3 b_0~ y^2+20 (c_2-2) k-20 k^2 ~y\right)+\frac{\Omega  e^{k y} \left(20 c_2~ k-y \left(3 b_0~ y+20 k^2\right)\right)}{36 c_2~ k}
\ee

As $\Omega\to 0$,we recover the solution obtained in \cite{Elahi}
and as both $\alpha ~\rm and ~\Omega \to 0$, \ref{A1-solution} reduces to the RS solution \cite{RS1}. We can perform a similar analysis for $\Omega<0$ considering the hyperbolic cosine function instead in the ansatz solution.

\subsection{Brane Tensions}
The field equations \ref{bulk-eqn} and \ref{boundary-eqn} are supplemented by the boundary conditions (up to leading order in $\alpha$)
\begin{equation}\label{boundary-condition}
\left[A^{\prime}(y)\right]_{i}=\frac{\epsilon_{i}}{12} \mathcal{V}_{i}
\end{equation}
where $\epsilon_{\mathrm{hid}}=-\epsilon_{\mathrm{vis}}=1$. We want to calculate the leading order corrections to the brane tensions in this setup. Using \ref{A1-solution} in \ref{ansatz}, we have the leading order correction to the warp factor. Substitute this expression in \ref{boundary-condition}, we obtain the following expressions for brane-tensions: 
\ba
\rm \mathcal{V}_{\text{hid}}=&&\rm \frac{12 k \left[1+\frac{\omega ^2}{\text{c2}^2}+\frac{10}{3} \alpha  k^2 (c_2-1) \left(1-\frac{\omega ^2}{c_2^2}\right)\right]}{\left(1-\frac{\omega ^2}{c_2^2}\right)+\frac{10}{3} \alpha  k^2 \left[\frac{\omega ^2}{c_2}+(c_2-2)\right]}\\
\rm \mathcal{V}_{vis}=&&\rm -\frac{12 k \Bigl\{1+\frac{\omega ^2 }{c_2^2}e^{2 \pi  k r_c}+\alpha  \left[\mathcal{B}(r_c)+\frac{\mathbf{G}(r_c) \omega ^2}{3 c_2^2} e^{2 \pi  k r_c}\right]\Bigr\}}{1-\frac{\omega ^2 }{c_2^2}e^{2 \pi  k r_c}+\alpha  \left[\frac{\mathcal{H}(r_c) \Omega  }{18 c_2^2 k}e^{2 \pi  k r_c}+\mathcal{J}(r_c)\right]}
\ea
where
\ba
    &&\mathcal{B}(r_c)=\pi  r_c \left(-\frac{1}{2} \pi  b_0 k r_c+b_0-\frac{10 k^3}{3}\right)+\frac{10}{3} (c_2-1) k^2\\
    &&\mathbf{G}(r_c)=\pi  r_c \left(\frac{3}{2} \pi  b_0 k r_c+3 b_0+10 \text{c2} k^3\right)-10 (c_2-1) k^2\\
    &&\mathcal{H}(r_c)=20 c_2 k-\pi  r_c \left(3 \pi  b_0 r_c+20 k^2\right)\\
    &&\mathcal{J}(r_c)=\frac{1}{6} k \left(-3 \pi ^2 b_0 r_c^2+20 (c_2-2) k-20 \pi  k^2 r_c\right)
\ea

In the limit $\alpha\to 0$,~
\ba
\rm \mathcal{V}_{\text{hid}}=12 k \LT \frac{1+\frac{\omega^2}{c_2^2}}{1-\frac{\omega^2}{c_2^2}}\RT; \hspace{5mm} \mathcal{V}_{\text{vis}}=-12 k \LT\frac{1+\frac{\omega^2}{c_2^2}e^{2k\pi r_c}}{1-\frac{\omega^2}{c_2^2}e^{2k\pi r_c}}\RT
\ea
consistent with the results of \cite{SD}.
In this case, $\cal{V}$$_{hid}$ is always positive and  $\cal{V}$$_{vis}$ is negative for the entire range of $\Omega>0$ for $\alpha$ and $b_0$ positive.

\section{Modulus Stabilization}
As has been discussed earlier modulus stabilization is an essential criterion for warped braneworld scenarios. Without a stable value for the moduli, no low-energy phenomenology can be extracted from the higher dimensional models. As a result, the study of the stabilization mechanism of higher dimensional models is of utmost importance to make the model viable and relevant.
The stabilized modulus field has wide phenomenological implications in the domain of cosmology and collider physics. Goldberger and Wise \cite{GW}-\cite{GW2} came up with a stabilization mechanism for the original Randall Sundrum model, which resolved the gauge hierarchy problem. In their work, a bulk scalar field was introduced which in turn generated a potential to stabilize the modulus field to the desired vev. However, they ignored the back reaction of the bulk scalar field onto the background geometry, which was later taken into account in \cite{Csaki}. Further, keeping in mind the scalar-tensor equivalence in $f(R)$ gravity theories \cite{defelice}-\cite{nojiri2}, efforts were made to achieve modulus stabilization with the conformally transformed metric\cite{ssg1}-\cite{ssg4}. In our previous work, we could achieve modulus stabilization in an original RS-like scenario in a $f(R)$ gravity background, which included higher curvature correction to the Einstein Hilbert action. On the other hand, it was shown that upon generalizing the original RS model to a non-flat braneworld scenario, with Einstein Hilbert action in the bulk, the modulus potential had a metastable minimum \cite{Banerjee}. In our present work, we are interested in examining the nature of the modulus potential in a generalized non-flat braneworld scenario in $f(R)$ gravity background. The vacuum expectation value of the modulus field, $T$, will yield the compactification radius, $r_c$, in our model.\newline 
\newline Considering only the bulk action from \ref{(1)}, after integrating out the extra dimension to obtain an effective 4-D theory, we will get a kinetic term for the modulus field, it's non-minimal coupling to the scalar curvature and its potential. The analytic expression for V(T) in terms of basic functions is not possible. Therefore, we have obtained a numerical solution for the potential for different values of the model parameters depicted by the plots below.

\begin{equation}
\label{(29)}  
\rm V(T) =\rm  2 \int_{0}^{\pi }{d\phi\,T\,e^{-4\text{A($\phi\, ,T$)}}\,\text{B($\phi\, ,T$)}\,(\text{R}+\alpha \text{R}^2-\Lambda_5)}\Bigg|_{\rm potential ~part}
\end{equation}

\begin{figure}[hbt!]
\begin{subfigure}{.450\linewidth}
  \includegraphics[width=\linewidth]{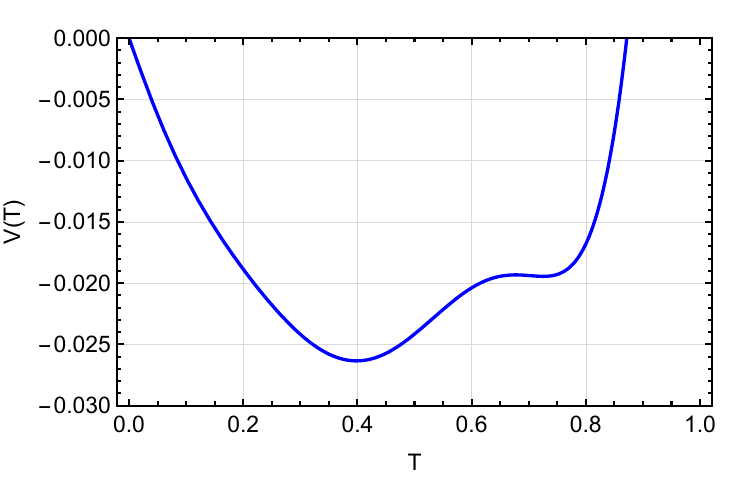}
  \caption{V(T) vs T for $\alpha=0.2$ and $b_{0}=10$.}
  \label{fig2}
\end{subfigure}\hfill 
\begin{subfigure}{.440\linewidth}
  \includegraphics[width=\linewidth]{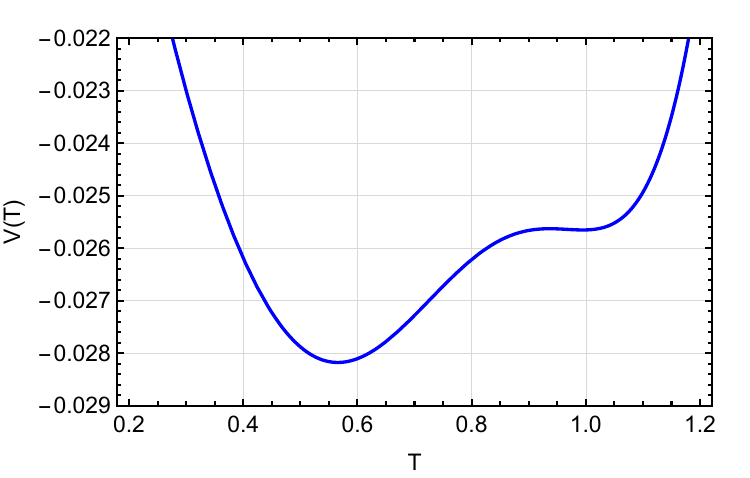}
  \caption{V(T) vs T for $\alpha=0.2$ and $b_{0}=5$}
  \label{fig3}
\end{subfigure}
\medskip 
\begin{subfigure}{.450\linewidth}
  \includegraphics[width=\linewidth]{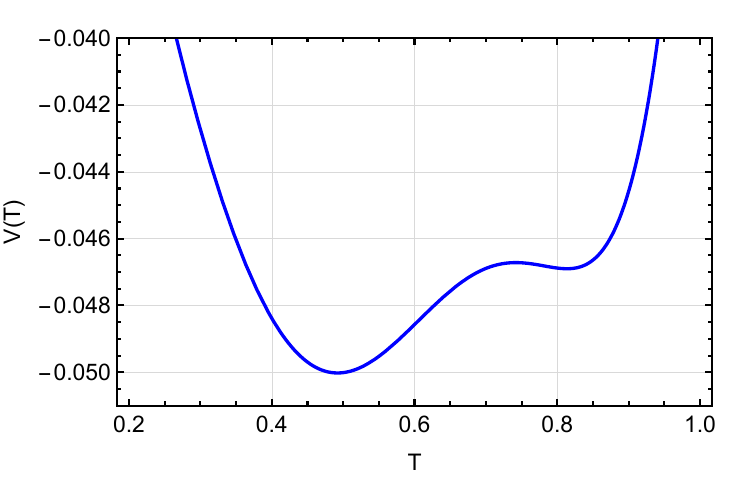}
  \caption{V(T) vs T for $\alpha=0.16$ and $b_{0}=10$}
  \label{fig4}
\end{subfigure}\hfill 
\begin{subfigure}{.440\linewidth}
  \includegraphics[width=\linewidth]{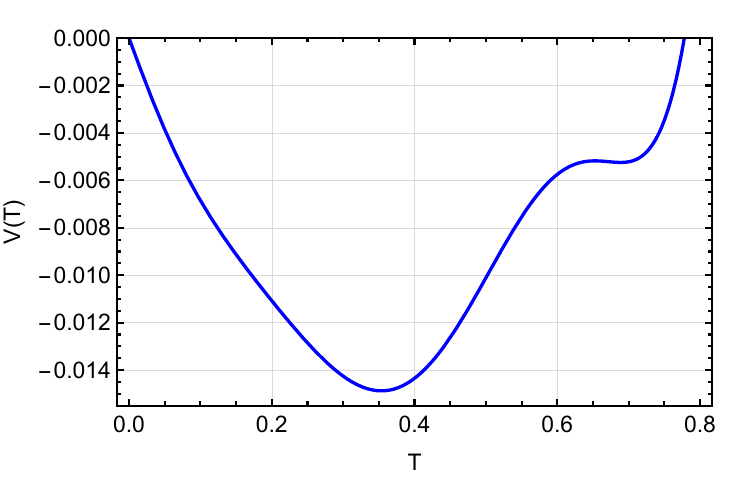}
  \caption{V(T) vs T for $\alpha=0.22$ and $b_{0}=10$}
  \label{fig5}
\end{subfigure}

\caption{Plots of modulus potential  for different values of $\alpha$ and $b_{0}$. }

\label{fig:potential}
\end{figure}
\noindent

    Stabilization can be achieved for $0<\alpha<1$, suggesting that our perturbative approach works. This stabilization criterion is the same as the one that we had obtained for the flat brane scenario \cite{Elahi}. As $\alpha \xrightarrow{}0$, only a metastable minimum occurs. This in turn further reinstates the fact that stable minima mainly arose due to the $\alpha$ correction to the warp factor, and as this correction term tends towards zero, the result is the same as the result shown in \cite {Banerjee}.\\
    To be consistent with $f'(R)>0$, we obtain the following bound on the model parameters : $\alpha>0$ and $b_{0}>0$. Hence, our model can be applied to the cosmology sector as $\alpha>0$. Therefore, we have  $0<\alpha< 1$ and $b_{0}>0$ to obtain modulus stabilisation in ghost-free quadratic $\text{f(R)}$ gravity.\\
We can further see that for a fixed value of $\alpha$, as the value of $b_{0}$ increases, the minima of the potential shifts and now the stabilization occurs for a smaller value of $T(x)$. In \ref{fig2}, stabilization occurs at $T(x)\approx 0.4$( for $b_0=10$ ), whereas in \ref{fig3}, stabilization occurs at $T(x)\approx 0.55$( for $b_0=5$ ), for the same value of $\alpha$ ( $\alpha$ = 0.2 ).For a fixed value of $b_{0}$, as the value of $\alpha$ increases, stabilization occurs for a smaller value of $T(x)$. For instance, in \ref{fig4}, the stabilization occurs at $T(x)\approx0.5$ ($\alpha=0.16$), while in \ref{fig5}, the stabilization occurs at $T(x)\approx0.35 $ ($\alpha=0.22$), for the same value of $b_0$ ( $b_0=10$ ).
\subsection{Swampland Conjecture} 
As string theory predicts the existence of extra dimensions for its consistency\cite{Palti:2019pca}, the connection between any effective higher dimensional model on low energy brane and a class of string vacuum is always worthwhile to explore\cite{Ooguri:2006in, Brennan:2017rbf}.
It is important to explore whether the proposed higher dimensional model satisfies the swampland conjecture and thereby can provide a consistent low-energy description of the vast string landscape. In this context, we focus our attention on the restrictions that the modulus potential needs
to satisfy to obey the swampland conjecture.
In our model, $\rm |\partial_T V(T)/V(T)|\sim $$\cal{O}$$(1)$ near the vacuum expectation value of the modulus field $T$. For $\alpha=0.2$ and $b_0=10$, $\rm |\partial_T V(T)/V(T)|\sim $$\cal{O}$$(1)$ for $0.35< T< 0.45$. Whereas, for $\alpha=0.2$ and $b_0=5$, $\rm |\partial_T V(T)/V(T)|\sim $$\cal{O}$$(1)$ for $0.4< T< 0.7$. It can be shown that for a certain choice of model parameters, $V(T)>0$. Thus the conditions imply that the model qualifies to be a consistent low-energy description of string landscape.
    
\section{Conclusion}
Branes with a non-zero cosmological constant are natural generalizations of the warped geometry model to explore the cosmological implications of braneworld models. Such a scenario was earlier shown to originate when bulk curvature contribution fails to cancel exactly the brane tension due to the presence of additional matter. The scenario with a de-Sitter brane was also shown to exhibit the presence of a metastable minimum for the modulus field which was absent for a pure RS-like scenario with flat branes. It was also exhibited that such a metastable minimum plays a crucial role in triggering a bouncing cosmological scenario leading to possible answers to avoid  cosmological singularity. The question however remained what happens to the moduli field once it crosses the metastable minimum and rolls down? 
In this work, we have demonstrated that the presence of a higher curvature term in the bulk leads to a stable minimum alongside a metastable minimum.
This eventually leads to an auto-modulus stabilization by the geometry itself and brings in interesting phenomenological consequences of a radion field with non-zero vev. The evolution of the scale factor in the presence of such a modulus potential with a non-minimal kinetic term for the modulus field may now lead to a 
bouncing scenario for an appropriate choice of the parameters. We have also verified that the swampland criteria are satisfied with the proper choice of model parameters and therefore the model proposes a consistent effective description of string landscape with viable ultraviolet description in the quantum regime.

\section*{Acknowledgement}
SGE and SSM would like to extend their gratitude to IACS where they carried out this work. 

\end{document}